\documentclass[reprint]{JASA}

\newcommand{\e}{\text{e}}
\newcommand{\jimag}{\text{j}}

\begin{document}

\title[JASA/Sample JASA Article]{Bayesian characterization of porous media using three-microphone tube method in extended frequency ranges}
\author{Ziqi Chen}
\email{chenz33@rpi.edu}
\author{Ning Xiang}
\affiliation{Graduate Program in Architectural Acoustics, School of Architecture,  Rensselaer Polytechnic Institute, Troy, NY 12180}





\preprint{Author, JASA}	

\date{\today}

\begin{abstract}
The characteristic impedance and the propagation coefficient are among the most important parameters for evaluating the acoustic performance of porous materials. This work investigates the influence of cylindrical modes in an impermeable tube and applies multiple microphones distributed along the tube circumference within the three-microphone framework to extend the valid frequency range of characteristic impedance measurement. During the extended broadband measurements, discontinuities or phase jumps are observed in the experimentally measured propagation coefficient of the porous material under test. A Bayesian inference is applied in a sequential manner to estimate the unwrapped propagation coefficient and characteristic impedance. The results demonstrate that the inferred parameters accurately capture the behavior of the transfer function, allowing accurate parameter estimation. 
\end{abstract}


\maketitle

\section{introduction}

The characteristic impedance and the propagation coefficient are among the most important parameters for evaluating the acoustic performance of porous materials. Impedance tube measurements are commonly used to characterize isotropic and homogeneous materials~\citep{salissou2010,song2000}. The three-microphone impedance tube method~\citep{salissou2010} represents one of the most widely adopted techniques in the acoustical materials community for experimentally characterizing these acoustic properties.

This work applies Bayesian inference to cope with challenges when extending the valid frequency range of tube measurements using averages of multiple microphones. 
During the extended broadband measurements, discontinuities or phase jumps are observed in the experimentally measured propagation coefficient of the porous material under test. This behavior arises because of the inverse cosine function, a necessary operation when determining the propagation coefficient of the porous materials. The inverse cosine function is value-ambiguous. 
While several approaches have been developed to address phase unwrapping problems~\citep{Tribolet1977}, the unwrapping of the $\arctan(\cdot)$ function is well established, yet fundamentally different from the unwrapping of the function $\arccos(\cdot)$  considered in this work. To the best of the authors' knowledge, the proposed unwrapping approach using Bayesian inference has not yet reported in the major acoustics literature. 

First, this work combines cylindrical mode decomposition~\citep{ABOM1989,cuenca2013angular} with the three-microphone transfer function method~\citep{salissou2010} to extend the upper limit frequency without reducing the tube diameter and material sizes. Multiple microphones are employed within the three-microphone framework to suppress circumferential modes at high frequencies in the impedance tube. 
This approach expands the valid frequency range of a tube with a diameter of 1.5~inches (38.1~cm) using the three-microphone method to about 9.5~kHz, nearly twice that of the conventional method.

Second, this work identifies a discontinuity in the estimated propagation coefficient and characteristic impedance at high frequencies. This discontinuity arises from the inherent limitations of the $\arccos(\cdot)$ function. 
It is nearly impossible to analytically recover the correct phase using the $\arccos(\cdot)$ function alone when only the transfer function—i.e., the cosine of the unknown, complex-valued phase function—is available.

To address this issue, this work applies Bayesian inference sequentially. Recently, \citet{eser2023free} has applied a similar approach in the free-field characterization of absorbing materials over a range of frequencies. However, this work essentially solves the phase discontinuity problem.   
Together, these contributions provide a robust framework for extending impedance tube measurements to higher frequencies while yielding stable, physically consistent propagation coefficients and characteristic impedances of the porous media. 

This paper is arranged as follows. Section~\ref{sec:method} discusses the cylindrical modes and the experimental setup for the three-microphone transfer function method by averaging multiple-microphones. Section~\ref{sec:phase} presents the phase discontinuity and parameter estimation by using Bayesian inference sequentially. Section~\ref{sec:discuss} discusses the application of sequential Bayesian inference in this work, and Section~\ref{sec:conc} concludes the paper. 

\section{Method}\label{sec:method}
The three-microphone transfer function approach~\citep{iwase1998new,salissou2010} has been widely used to measure the characteristic impedance and the propagation coefficient of fluid-equivalent materials. However, the assumption that only plane waves propagate within the tube restricts the usable measurement frequency range.
In this work, a multiple-microphone strategy is incorporated into the three-microphone transfer function approach to extend the valid frequency range. By averaging the impulse responses measured at each microphone position within one cross-sectional plane, the circumferential modes are effectively canceled.
\subsection{Cyclindrical harmonics decomposition}


Given sound waves propagating in a circular tube, wave equations in cylindrical coordinates govern the sound wave propagation.
Separating variables, the solution of the wave equation of the sound pressure $p$ at a certain location has the general form
\begin{equation}
    p(\phi,r,z) = \Phi(\phi)R(r)Z(z),
\end{equation}
where $\Phi(\phi)$ represents an angular component, also termed a circumferential function, $R(r)$ is the radial component, while $Z(z)$ is the solution component along the length of the tube. The general solutions for the sound pressure $p$ along the radial coordinates are Bessel functions \citep{rienstra2015fundamentals}
\begin{equation}
    p(r) = R_{mn}(r) =  J_m\left(\alpha_{mn}\cdot\frac{r}{a}\right),
    \label{eq: boundary r=a}
\end{equation}
where $J_m$ is the Bessel function of $m$th order, $m$ is the circumferential modal number, $n$ is the radial modal number, $\alpha_{mn}$ is the radial and axial wave numbers. With hard tube walls, the boundary condition at $r = a$ is that the radial velocity $v_r = 0$ \citep{rienstra2015fundamentals}. This implies that $\alpha_{mn}$ follows the boundary condition
\begin{equation}
     \left.\frac{\partial p}{\partial r}\right\vert_{r = a} = 0.
\end{equation}
Therefore, $\alpha_{mn}$ are equal to the non-negative zeros $q_{mn}$ of the derivative of the Bessel function $J'_m$, written as
\begin{equation}
    J'_m\left(q_{mn}\right) = 0.
\end{equation}
For each zero, one can calculate the cut-off frequency for different modes
\begin{equation}
    f_{mn} = \dfrac{q'_{mn} c}{d},
    \label{eq:fmn}
\end{equation}
where $q'_{mn}$ is the modal factor, $q'_{mn} = q_{mn}/\pi$, $c$ is the sound speed in the tube, and $d$ is the diameter of the tube. In general, a mode propagates or decays depending on the frequency being lower or higher than the cut-off frequency \citep{rienstra2015fundamentals}.
Table~\ref{tab:qmn} lists modal factors $q'_{mn}$ in Equation \eqref{eq:fmn}.

\begin{table}[ht]
\caption{ Modal factors $q'_{mn}$ in Equation \eqref{eq:fmn}.}\label{tab:qmn}
\begin{ruledtabular}
\begin{tabular}{c|cccc}
$q'_{mn}$       & $m=0$    & $m=1$    & $m=2$    & $m=3$   \\
\hline
$n=1$           & 0         & 0.586  & 0.972  & 1.337 \\
$n=2$           & 1.220     & 1.697  & 2.135  & 2.551 \\
\end{tabular}
\end{ruledtabular}
\end{table}

For a circular tube, the general solution of the circumferential function is
\begin{equation}
    \Phi(\phi) = \cos(m\phi).
\end{equation}
The microphones should be mounted flush with the interior surface of the tube. Therefore, the radial distances $r_0 = a$ are identical for each microphone at the same longitudinal position $z_0$. In other words, the radius $r$ and length $z$ at which the measurements are performed are constant. When  $N$ microphones are evenly spaced along the circumference of the tube, the average sound pressure $\overline p$ of $N$ microphones becomes
\begin{equation}
    \overline p(\phi,r=a,z=z_0) = \sum_m\sum_n \overline \Phi_m(\phi)R_{mn}(a)Z_{mn}(z_0),
\end{equation}
where $ \overline \Phi_m(\phi)$ represents the average circumferential function where microphones are assumed to be evenly distributed along the circumference\citep{panzer2019extracting}
\begin{equation}
     \overline \Phi_m(\phi) = \frac{1}{N}\sum_{k=0}^{N-1}\cos(m\phi + m\cdot\frac{2k\pi}{N}).
    \label{eq: circum average}
\end{equation}
Solving Equation \eqref{eq: circum average} leads to
\begin{equation}
 \overline \Phi_m(\phi) = \begin{cases}
1, &m=0,\\
0, &m>0 \cap m<N.
\end{cases}
\label{eq:modes cancellation}
\end{equation}
\begin{figure}[h]
    \centering
    \includegraphics[width=0.95\linewidth]{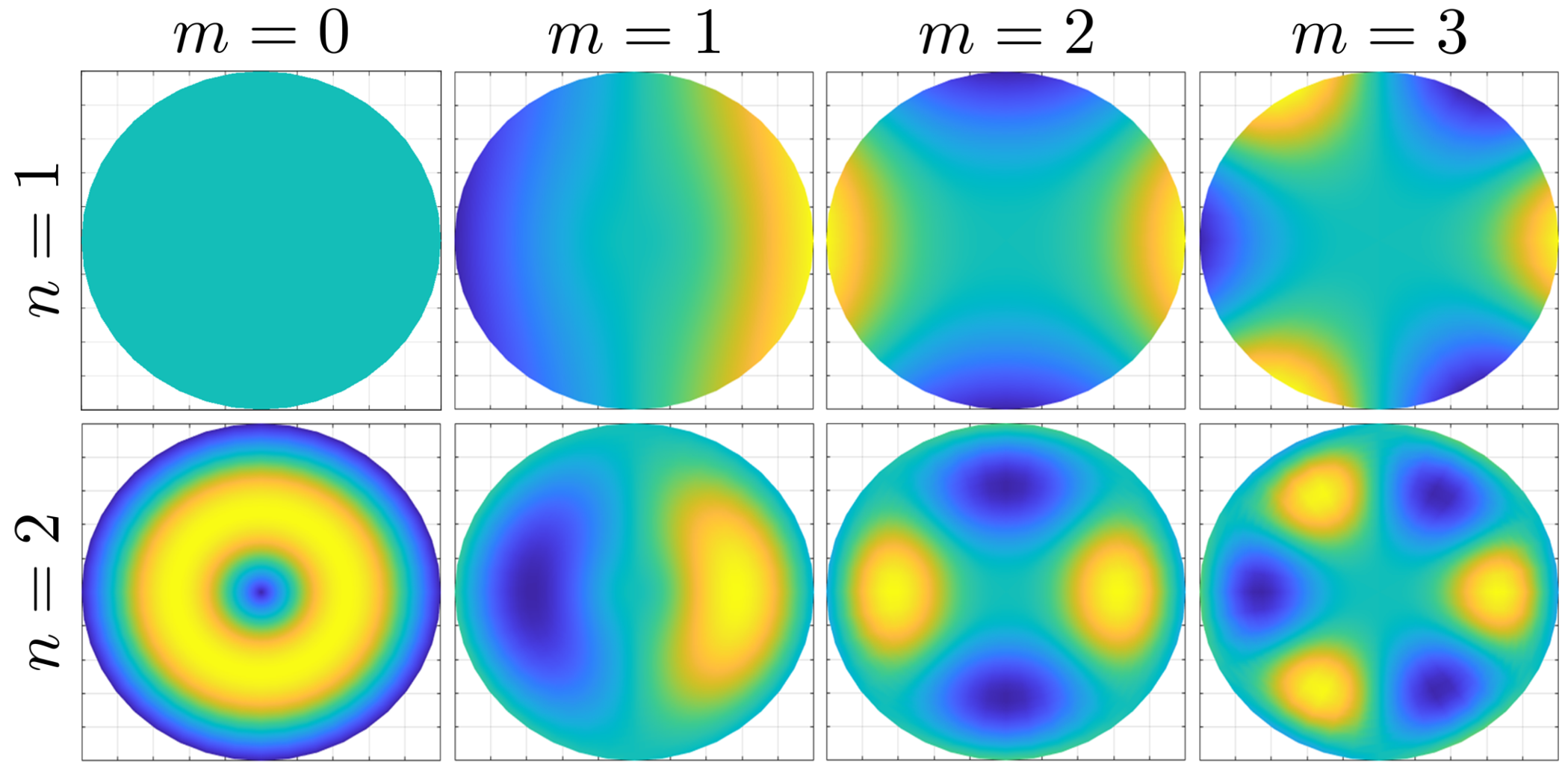}
    \caption{Cross-section of acoustic cylindrical modes in a circular tube.}
    \label{fig:cyc mode}
\end{figure}
Figure~\ref{fig:cyc mode} shows the modal distributions in the cross section of a circular duct. As illustrated, circumferential modes exhibit angular symmetry about the duct axis. By uniformly sampling $N$ angular positions on a circle of constant radius centered on the duct axis and averaging the measured sound pressures, all circumferential modes of order less than or equal to $N$ are effectively canceled.
Equation \eqref{eq:modes cancellation} indicates that using the average sound pressure of $N$ microphones would cancel the circumferential modes lower than the number $N$ of microphones along the circumference. 

However, radial modes still exist and cannot be suppressed, because microphones only measure the superposition of all modes. According to Equation \eqref{eq:fmn} and Table \ref{tab:qmn}, the cutoff frequency of the (0,2) mode is higher than that of the (2,1) mode and lower than that of the (3,1) mode. At any frequency higher than $f_{02}$, the radial mode is imposed on all the measured impulse responses. It cannot be canceled by averaging the signals of multiple microphones along the circumference. To achieve the maximum frequency range in a multiple-microphone measurement, one can embed three microphones evenly along the circumference of the circular tube wall. If more than three microphones are deployed along the circumference, the frequency range of multiple-microphone measurements is still limited by the cutoff frequency of the (0,2) mode.

\begin{figure}[ht]
    \centering
    \includegraphics[width=0.95\linewidth]{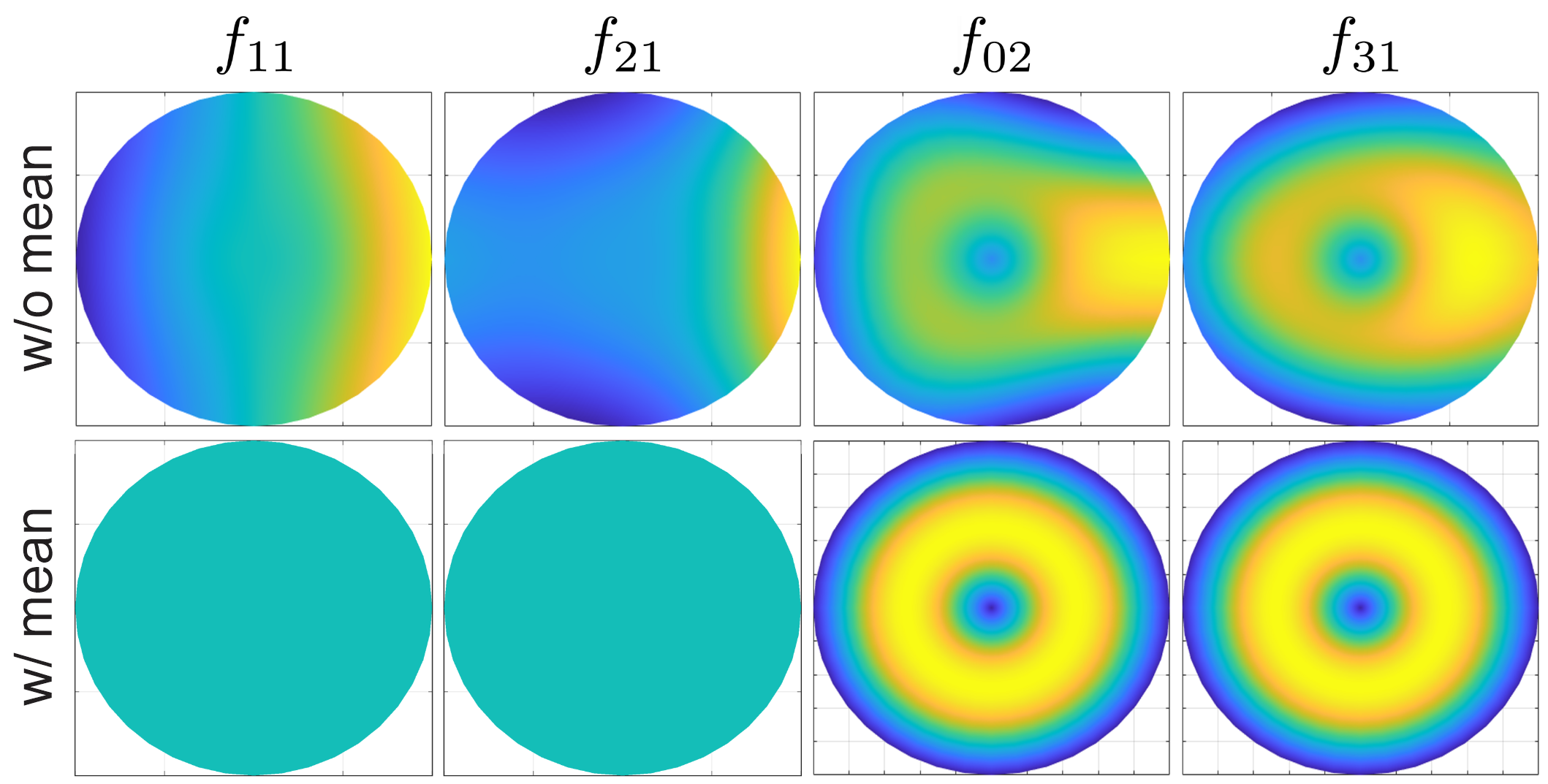}
    \caption{Cross-section of summed acoustic cylindrical modes in a circular tube without and with the multiple microphone method.}
    \label{fig:mode cancel}
\end{figure}

Figure \ref{fig:mode cancel} shows modal distributions over the cross-section of the superimposed acoustic fields in a circular tube with enough microphones. With multiple microphones evenly distributed along the circumference of the tube, higher circumferential modes would cancel. By doing this, only the fundamental plane wave would be left, which can be used to calculate the acoustic properties at high frequencies. There is an error above $f_{02}$ because of uncancelable radial modes.

\subsection{Multiple-microphone three-mic transfer function method}
\begin{figure}[ht]
    \centerline{
    \includegraphics[width=\linewidth]{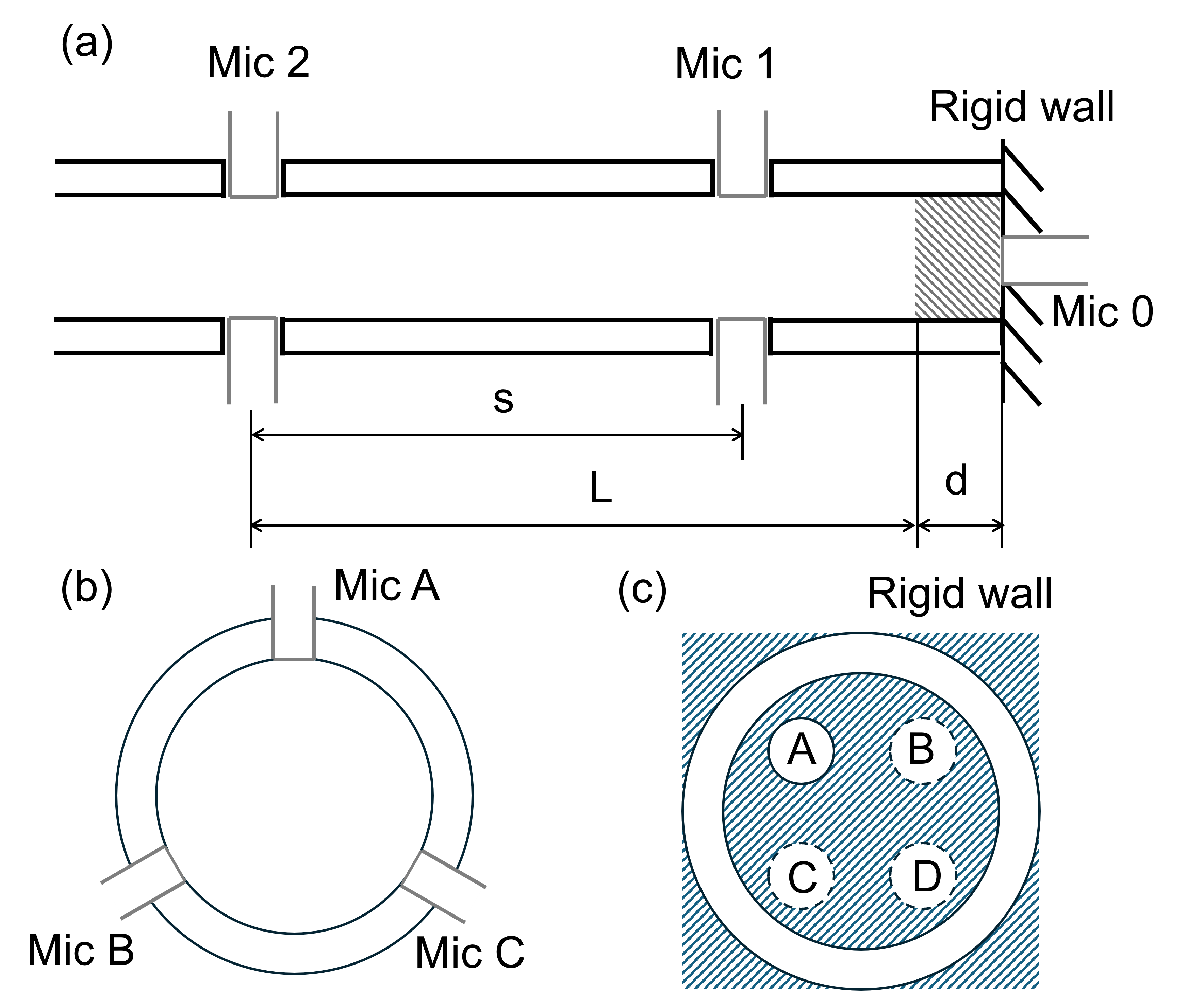}}
    \caption{Setup for Multiple-microphone three-microphone transfer function measurement. (a) General configuration for the impedance tube. (b) Multiple microphones distributed evenly along the circumference at one cross-sectional plane in front of the material (for Mic 1\&2). (c) Four hypotheitic  microphone positions at the back surface of the material (for Mic 0) through rotating the rigid backing four times.}
    \label{fig:MMP}
\end{figure}

The three-microphone transfer function method \citep{salissou2010} is a widely used method for measuring characteristic impedance. Figure \ref{fig:MMP} shows a sketch of the impedance tube setup for the three-microphone transfer function method. However, the sound wave modes inside the tube limit the traditional three-microphone method~\citep{ASTM1050}. The transfer function method assumes that the sound wave propagating along a tube is a plane wave. The statement is generally true at low frequencies. As the frequency increases, the circumferential and radial modes will be superimposed on the acoustic field. To address the limitations of traditional impedance tube measurements, Sanada \citep{sanada2018extension} and Panzer \citep{panzer2019extracting} employ an averaging method of multiple microphones based on cylindrical mode decomposition. With multiple microphones evenly spaced around the circumference within a cross-sectional plane, the mean signal from multiple microphones could cancel the circumferential modes at high frequencies. Figure \ref{fig:MMP}(b) shows the cross-section of the impedance tube with multiple microphones along the circumference.
This work applies the multiple-microphone method to the three-microphone transfer function method \citep{salissou2010} to extend the frequency range of characteristic impedance measurement. In addition to the multiple microphones along the circumference, four microphones are placed center-symmetrically at the back surface of the material (Figure \ref{fig:MMP}(c)). As mentioned in the previous section, the upper limit frequency for the measurement with either three or four microphones along the circumference is the cutoff frequency of (0,2) mode. 

Note that the room modes inside a square tube would be different from the cylindrical modes discussed in this work \citep{yang2025data}. The cutoff frequency would be different in the square tube because the modes differ.

With circumferential modes cancelled, the transfer function $\underline H_{\,ij}$ between two averaged pressure spectra measured at two microphone positions is
\begin{equation}
\underline{H}_{\,ij} =\dfrac{\,\tilde{\underline{P}}_{\,i}}{\,\tilde{\underline{P}}_{\,j}\,}, \label{EQ:averTrs}
\end{equation}
where $\tilde{\underline{P}}_{\,i}$ and $\tilde{\underline{P}}_{\,j}$ represent the average spectra of the sound pressure signals measured with the multiple microphones along the circumference at microphone positions $i$ and $j$, $\tilde{\underline{P}}_{\,i} = ( \sum_{n=1}^N \underline{P}_{\,i,n})/N$. 
Either the mean value of the temporal impulse responses before the Fourier transform or the mean value of the frequency responses after the Fourier transform serves the purpose. 

As illustrated in Fig.~\ref{fig:MMP}, the surface reflectance $\underline R$ and the surface impedance $\underline Z_s$ of the testing material are determined~\citep{Chung1980a} by 
\begin{equation}
    \underline{R} = \dfrac{\,\underline{H}_{12} - \e^{-\underline{\gamma}_0 s}\,}{\e^{\,\underline{\gamma}_0 s} - \underline{H}_{12}} \, \e^{2 \underline{\gamma}_0 L},
    \label{eq:surface_reflectance}
\end{equation}
and
\begin{equation}
    \underline Z_s = Z_0\dfrac{\,1+\underline R\,}{\,1 - \underline R\,},
        \label{eq:surface_impedance}
\end{equation}
where $\underline{\gamma}_0$ is the propagation coefficient of air in the tube, $s$ is the separation between the microphones Mic~1 and Mic~2, $L$ is the distance from the front surface of the porous material to Mic 2, and $Z_0$ is the characteristic impedance of air. The key to obtaining the propagation coefficient and the characteristic impedance of the porous material under test is to calculate the transfer function $H_{d0}$, being the spectral ratio of the sound pressures at the front surface of the material and at the back surface of the material \citep{iwase1998new}. It is given by
\begin{equation}
    \underline H_{\,d0} = \frac{1 + \underline{R}}{\e^{\,\underline{\gamma}_0(L-s)} + \underline{R}\,\e^{-\underline{\gamma}_0(L-s)}} \, \underline{H}_{10},
    \label{eq:H_d0}
\end{equation}
where $ \underline{H}_{10}$ represents the transfer function between the microphone Mic~1 and the Mic~0 (see Fig.~\ref{fig:MMP}).
With the surface reflectance $\underline R$ and surface impedance $\underline Z_s$ of the sample being given in Eq.~\eqref{eq:surface_reflectance} and ~\eqref{eq:surface_impedance}, the transfer matrix theory~\citep{Xiang2021} yields the complex propagation coefficient $\underline k$ and the characteristic impedance $\underline Z_c$ of the porous sample under test~\citep{salissou2010} as follows
\begin{equation}
\begin{aligned}
        \underline{\theta} = \underline{k}\,d &=  \cos^{-1}\underline H_{d0}\\
                &=  \cos^{-1}\left[ \frac{1 + \underline{R}}{\e^{\,\underline{\gamma}_0(L-s)} + \underline{R}\,\e^{-\underline{\gamma}_0(L-s)}} \, \underline{H}_{10} \right],
    \label{eq:k}
\end{aligned}
\end{equation}
and
\begin{equation}
    \underline{Z}_c = \jimag \underline Z_s \tan(\underline{k}\, d),
    \label{eq:Zc}
\end{equation}
where $d$ is the thickness of the porous material under test. 

Equation~\eqref{eq:Zc} explicitly indicates that with the experimentally measured surface impedance $\underline{Z}_s$ via Eq.~\eqref{eq:surface_impedance} and the material thickness $d$ being avaiable, the characteristic impedance $\underline{Z}_c$ of the porous material under test is only obtainable when the propogation coefficient $\underline k$ of the porous material is experimentally determined via Eq.~\eqref{eq:k}.

\section{Phase Unwrapping Using Bayesian Inference}\label{sec:phase}

\begin{figure}[ht]
    \centering
    \includegraphics[width= 0.4\linewidth]{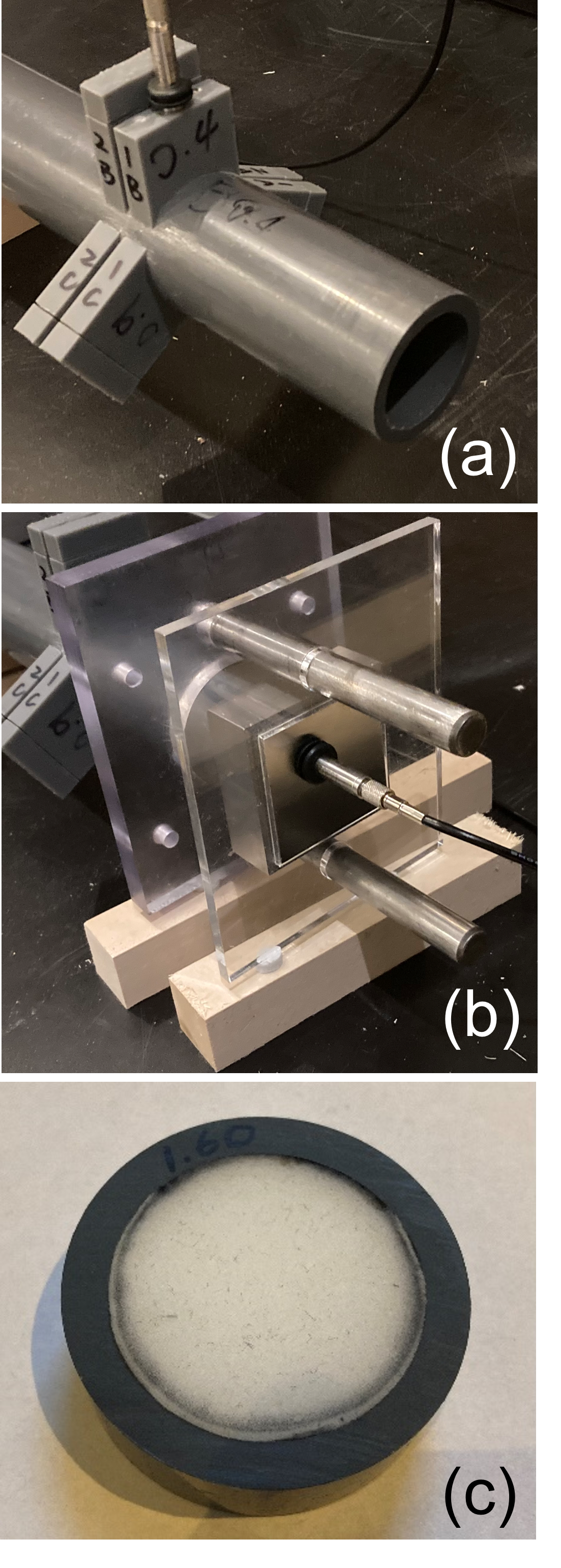}
    \caption{Setup for the multiple-microphone characteristic impedance measurement for (a) Mic 1\&2; (b) Mic 0; (c) the testing material.
    }
    \label{fig:Tube}
\end{figure}

Experimental measurements are carried out following the standard transfer function method~\citep{ASTM1050,ISO10534}. The propagation coefficient $\underline \gamma_0$ of air in the tube is calibrated following the dissipation estimation method proposed by Chen and Xiang \citep{chen2024bayesian}. 
Figure~\ref{fig:Tube} shows the photos of the measurement setup used in this work. The tube is made of PVC, 2~m long and 6.4~mm thick (0.25 inch), with an inner tube diameter of 37.5~mm (1.48 inches), which would dictate an upper limit frequency of 5.4~kHz without averages of multiple microphone signals, 
The rigid termination is established using a solid metal block. The first microphone position, denoted as microphone position 1 in Figure \ref{fig:MMP}(a), is located 10.2 cm (4 inches) from the front surface of the material under test, and microphone position 2 is located 11.4~cm  (4.5 inches) away from the front surface of the material under test, ensuring a microphone separation of $s = 1.2$~cm. Such a small microphone position favors high frequency validity towards 9.5~kHz, it is still suitable to obtain accurate measurements from 2.5~kHz upwards~\citep{ISO10534}. Microphone position 0 is located 1~cm away from the axial center of the rigid metal block. During sound pressure measurements at microphone position~0, the metal block is rotated by 90 degrees four times to obtain impulse responses from all four microphone positions flush at the rigid backing, as illustrated in Fig.~\ref{fig:MMP}(c).

The impulse responses at all microphone positions in the tube are experimentally measured in a sequential manner suggested by Chu \citep{chu1986} to avoid microphone phase mismatches. The impulse responses are windowed to eliminate all unwanted reflections except the direct sound and the direct reflection from the testing samples. The three signals measured within each cross-sectional plane at microphone positions~1 and~2 are averaged, and the four signals measured at microphone position~0 are also averaged. The mean windowed impulse responses at each microphone position are used to calculate the surface reflectance, the propagation coefficient, and the characteristic impedance of the material.

The material under test is melamine foam, mounted within one segment of the tube and attached to the tube end (Fig.~\ref{fig:Tube}(c)), with the interface sealed with petroleum jelly (Vaseline) to prevent possible air leakage.
To measure distinct sound pressure impulse responses at microphone positions, a 1/4 inch microphone (PCB, Inc.) is introduced into the tube, as shown in Fig.~\ref{fig:Tube}, following the methodology outlined by Chen and Xiang~\citep{chen2022bayesian}. 

\subsection{Phase Discontinuity}
Within an achievable frequency range between 2.5~kHz and 9.5 kHz, one challenge in acoustic measurements often observed at high frequencies is the discontinuous, wrapped propagation coefficient. The propagation coefficient $\underline k$ in Equation \eqref{eq:k} is proportional to the inverse cosine of the transfer function 
\begin{equation}
\underline H_{\,d0} = \cos (\,\underline k \,d).
    \label{eq:cosine}
\end{equation}
The unknown phase function is the propagation coefficient times the material's thickness: $\underline \theta = \underline k d$, being a complex-valued phase function of frequency, of which the inverse cosine function is multivalued 
\begin{eqnarray}
    \cos^{-1}\left( \underline H_{\,d0} \right) = -\jimag\log\left(\underline H_{\,d0} \pm \jimag\sqrt{1 - \underline H_{\,d0}^2}\right).
    \label{eq:acos}
\end{eqnarray}
As indicated by the square-root term in Equation~\eqref{eq:acos}, the evaluation of the inverse cosine function involves the sine of the unknown, complex-valued phase function $\underline{\theta}$. The absence of information regarding the correct sign ($+/-$) of the sine function leads to errors in the inversion of the cosine, which manifest as phase discontinuities. This issue presents a unique challenge in the experimental characterization of porous media using the three-microphone method, particualrly over such an extended frequency range up to 9.5~kHz.

\begin{figure}
    \centering
    \includegraphics[width=0.8\linewidth]{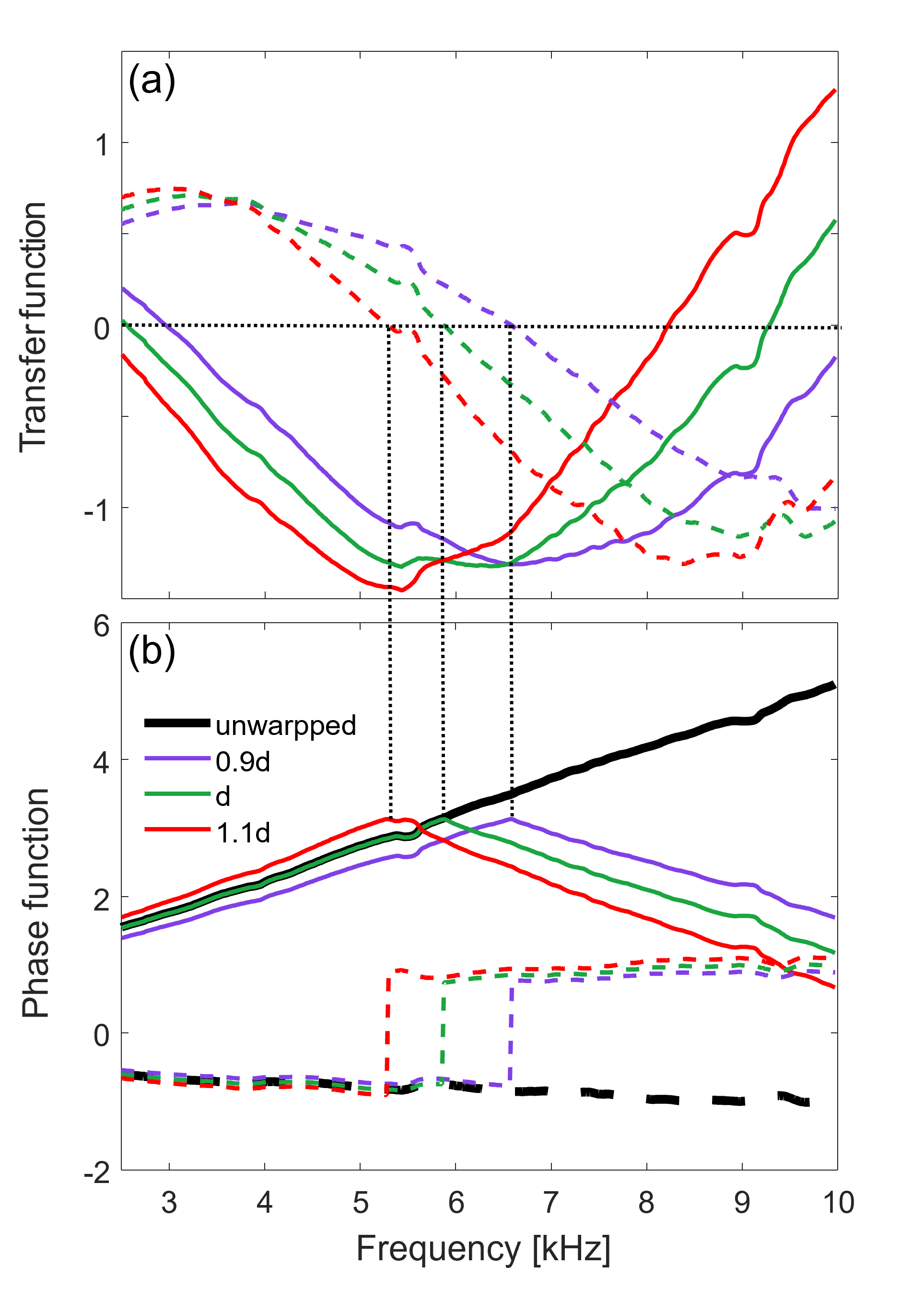}
    \caption{Transfer functions and calculated propagation coefficients for the hypothetical porous material of three different thicknesses, $d = 2.75$~cm. (a) Transfer functions; (b) Propagation coefficients. }
    \label{fig:ThkScale}
\end{figure}
Figure \ref{fig:ThkScale} illustrates the transfer functions $\underline H_{d0}$ and associated propagation coefficients for different thicknesses of mateirals, based on the experimental measurement of porous material of one thickness ($d$ = 2.75 cm). Given the experimental data  $\underline H_{d0}$, the propagation coefficient $\underline k$ is first calculated via inversion of Eq. (16) and unwrapped, then multiplying of two different thicknesses $d$ to obtain the phase function $\underline \theta = \underline k d$ and the corresponding transfer function $\underline H_{d0} = \cos(\underline k\, d)$, followed by inversion of Eq.~\eqref{eq:cosine}.

The black solid and dashed lines in Fig.~\ref{fig:ThkScale}~(b) are the real parts and imaginary parts of the unwrapped propagation coefficient, respectively. In the figure, the frequency of the phase jump increases with decreasing material thickness. As the material thickness increases, the phase jump occurs at a lower frequency. When the imaginary part of the transfer function $\underline H_{d0}$ crosses zero as in Fig.~\ref{fig:ThkScale}~(a), the wrapping will occur in the associated propagation coefficient. 
The difference in the wrapping behavior of the real and imaginary parts arises from the properties of the $\arctan(\cdot)$ function. In this specific set of experimental results, the $\arctan(\cdot)$ function shifts the real part to $-\pi$ when it reaches $\pi$ at approximately 6~kHz, whereas the $\arccos(\cdot)$ function changes the sign of both the real and imaginary parts. Consequently, the wrapping behavior observed in the real part differs from that observed in the imaginary part.

The frequency at which the phase jump occurs also varies with temperature and humidity. Changes in environmental factors, such as temperature, affect visco-thermal dissipations in porous materials \citep{JKD1987, CA1992}, leading to changes in the propagation coefficient and characteristic impedance. Therefore, the propagation coefficient will reach the 'threshold' at different frequencies, resulting in different locations along the frequency axis where the phase jump occurs. Phase wrapping does not occur at relatively low frequencies; therefore, the propagation coefficient calculated at low frequencies can be reliably used to assign a prior distribution for the initial inference step, as elaborated in the following Section. 

According to Eqs.~\eqref{eq:k} and~\eqref{eq:Zc}, the calculation of both the propagation coefficient and the characteristic impedance relies on the complex phase function $\underline \theta = \underline k d$. Therefore, the wrapped phase function $\underline \theta$ results in a wrapped propagation coefficient and the characteristic impedance. To unwrap the complex-valued function $\underline \theta$, this work sequentially applies Bayesian parameter estimation in Section \ref{subsex:SeqBay}.

\subsection{Bayesian inference in a sequential manner}\label{subsex:SeqBay}
The solution to the wrapped-data problem is to apply Bayesian inference sequentially at each data point (frequency bin). This work treats each data point in $\underline{H}_{d0}$ independently as an individual observation. The mean of the estimated posterior at data point $\underline{\theta}_n$ then informs the mean of the prior distribution for the subsequent data point $\underline{\theta}_{n+1}$~\citep{eser2023free}. The posterior and the prior distributions of the unknown, complex-valued phase function $\underline{\theta}$ satisfy Bayes' theorem.
\begin{equation}
    \overbrace{P(\underline \theta|\underline H_{d0},\underline H_{M},I)}^\text{posterior} \propto 
    \overbrace{P(\underline H_{d0}|\underline \theta,\underline H_{M},I)}^\text{likelihood}\times 
    \overbrace{P(\underline \theta|\underline H_{M},I)}^\text{prior}, \label{eq:bayes}
\end{equation}
where $I$ is the background information, $\underline H_{M}$ represents the prediction model 
\begin{equation}
\underline H_{M} = \cos \underline \theta \approx \cos (\underline{k} \,d), \label{eq:model}    
\end{equation} 
and the transfer function $\underline H_{d0}$ represents the experimental data, and $(\underline{k} \,d)$ is the estimate of $ \underline \theta$.  The posterior is proportional to the multiplication of the prior and the likelihood. The background information $I$ includes that {\it the cosine function of the wave number predicts the transfer function $\underline H_{d0}$ at low frequencies}. It also includes that {\it the propagation coefficient should be continuous in the frequency domain of interest} and that {\it the width of the prior distribution is finitely constrained because the cosine function is periodic}.

The prior probability $P(\underline \theta|\underline H_{M},I)$ represents our prior knowledge of the complex phase value $\underline{\theta}$. The phase function calculated from Equation~\eqref{eq:k} is well limited below the 'wrapping threshold' at low frequencies (below, for example, 3~kHz) and is expected to vary smoothly within this frequency range. Since the complex phase function $\underline{\theta}$ is continuous, the values at adjacent data points should be close to one another. This continuity constraint is therefore encoded into the prior assignment for each subsequent data point. The prior is physics-informed rather than heuristic.

With only a weak degree of prior information, the initial prior for $\underline{\theta}_1$ is assigned to a uniform distribution according to the principle of maximum entropy~\citep{xiang2020}, centered around the initial phase value $\underline{\theta}_0$, with a width of $\pi$ for both the real and imaginary parts. Then the priors for the later data points $\underline{\theta}_{n\geq2}$ are assigned Gaussian distributions with the mean and variance of the posterior distribution of the preceding data points. The assignment of Gaussian distributions is based on the learned knowledge of the parameter values and ranges via the posterior distribution of the previous data point ($n=1$). In this way, a more informative prior with a narrower region is used to estimate the unwrapped phase function at the subsequent points along the frequency axis. 
The width of the initial prior cannot be too broad, since the cosine function is periodic. An excessively broad prior distribution may introduce ambiguities and lead to errors in the posterior distribution estimation.

The likelihood $P(\underline H_{d0}|\underline \theta,\underline H_{M},I)$ represents the degree of belief in the data given the hypothesis $\underline \theta$. The residual error between the data and the hypothesis determines the likelihood function. The residual error $\epsilon_n$ in the frequency domain at the data point $n$ is the square root of the sum of the squares of the real part and the imaginary part
\begin{equation}
        \epsilon_n^2  =  \mathrm{Re}^2(\underline H_{d0,n} - \underline H_{\mathrm{M},n}) 
         +  \mathrm{Im}^2(\underline H_{d0,n} - \underline H_{\mathrm{M},n}).
    \label{eq:error}
\end{equation}
According to the principle of maximum entropy~\citep{xiang2020}, all the prior knowledge should be encoded into the prior assignment. Application of the principle of maximum entropy leads to a Gaussian distribution assignment for the likelihood function. Due to the missing information on the variance within the Gaussian assignment, the variance of the likelihood function is marginalized to obtain a Student-t distribution
\begin{equation}
    p(\underline H_{d0,n}| \underline\theta_n,\underline H_\mathrm{M,n},I) = \dfrac{\Gamma(K/2)}{2}\left(\pi \epsilon_n^2 \right)^{-K/2},
    \label{eq:student}
\end{equation}
where $K$ is the total number of the data points, and $\Gamma(\,\dots)$ is the standard Gamma function. 

The prior assignment in this procedure can be divided into two cases: (1) at the first data point, the prior probability is assigned as a uniform distribution; and (2) at data point $n$, the prior probability is assigned as a Gaussian distribution whose mean and variance are assigned by the estimated posterior at data point $n-1$. The variance of this Gaussian prior is chosen to be slightly larger than that of the previous posterior to cover a sufficiently wide parameter range~\citep{eser2023free}. Thus, uniform sampling is used for the first data point, followed by Importance sampling~\citep{kloek1978importancesampling} with the assigned Gaussian prior as the weighting function for all subsequent data points. In this way, information from the posterior distribution is propagated into the prior knowledge at the next data point, thereby improving both the efficiency and accuracy of the estimation during the sampling process. The width of the prior distribution must be selected carefully, as the cosine function is periodic. An overly broad prior may lead to inaccurate estimates, since different values of $\underline{\theta}$ can correspond to the same associated $\underline{H}_\mathrm{M}$. 

\subsection{Estimation results}

\begin{figure}[ht]
    \centering
    \includegraphics[width=0.5\linewidth]{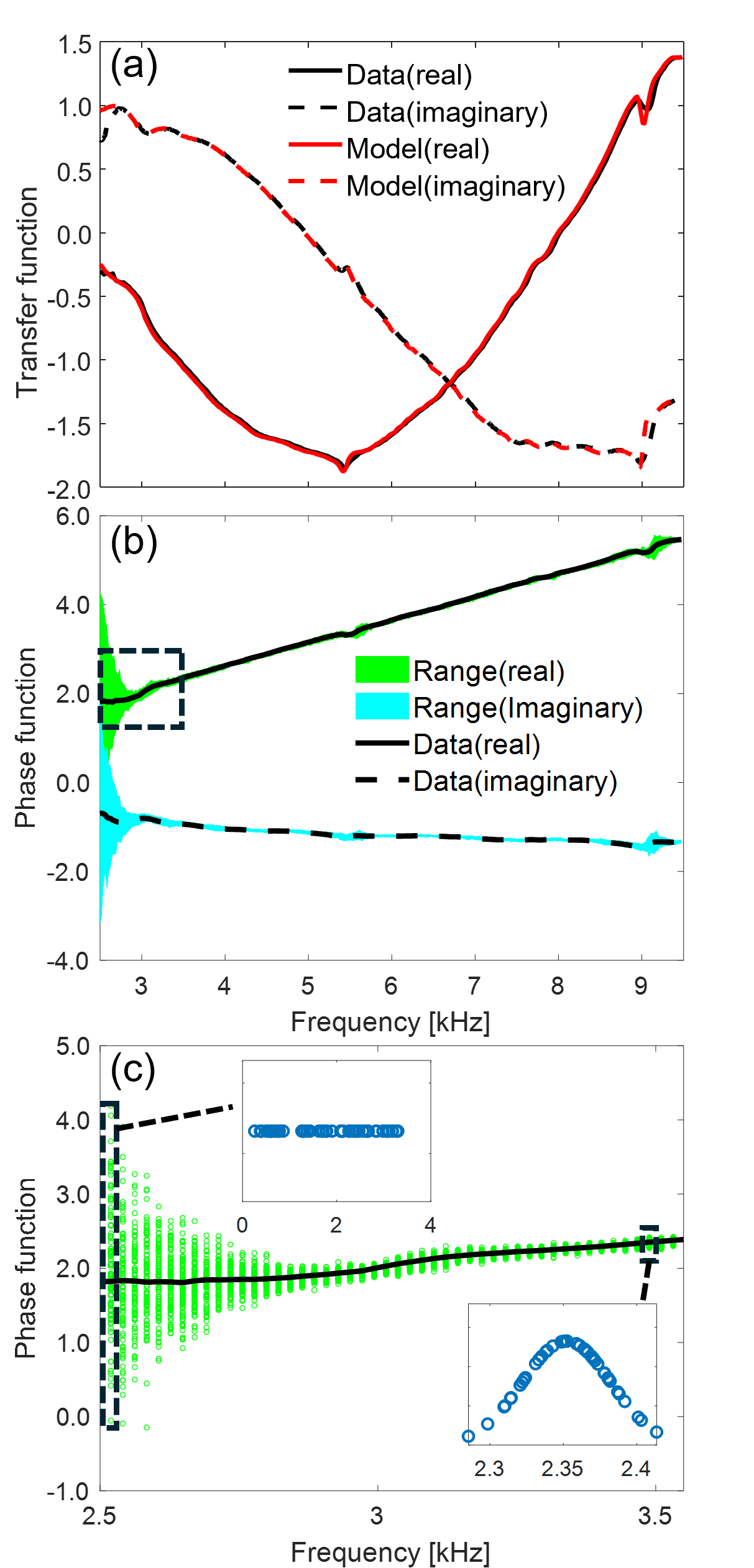}
    \caption{The unwrapped phase function and transfer function. (a) The estimated and the measured transfer function; (b) the estimated phase function; (c) the real part of the phase function from 2.5 to 3.5 kHz. The dotted blue points are the scatter plot of the prior distribution at 2.5 kHz and 3.5 kHz. }
    \label{fig:Seq}
\end{figure}

As discussed in Section~\ref{subsex:SeqBay}, Bayesian inference is applied sequentially to estimate the unwrapped propagation coefficient and the characteristic impedance. 
The achievable frequency range of interest within this work spans 2.5--9.5~kHz after averaging multiple microphone signals within cross-sectional planes. For the initial step, the parameter range is set to $\pi$ around the value obtained from the inverse cosine of the transfer function $\underline{H}_{d0}$ at 2.5~kHz. At each data point, 500 samples of a uniform/Gaussian distribution are used. The assignment of a uniform prior distribution around the data value in the initial step is consistent with the principle of maximum entropy~\citep{xiang2020} that no subjective preference to any values of the propagation coefficient is injected into the analysis. Based on the likelihood values of these samples, the posterior mean and variance are estimated and subsequently used to update the prior distribution for the next data point. To cover sufficient value ranges, the standard deviation of the prior distribution at each step is set to 1.2 times the posterior standard deviation from the previous data point.

Figure~\ref{fig:Seq}~(a) presents the transfer function measured experimentally. The solid and dashed red curves represent the modeled results inferred from the Bayesian framework, overlaid with the measured data shown by the black curves. The close overlap between the modeled curves and the measurements across the frequency range indicates good agreement, demonstrating that the inferred parameters accurately capture the behavior of the transfer function. 

Figure~\ref{fig:Seq} (b) illustrates the corresponding estimated unwrapped phase function $\underline{\theta}$ along with its associated sample ranges. The shaded regions indicate the parameter ranges spanned by the random samples, while the black solid line represents the mean value of the sampled posterior distribution. An inset highlights a zoomed-in frequency region, showing the tight clustering of samples and the smooth evolution of the unwrapped phase.

Figure~\ref{fig:Seq} (c) illustrates a portion of the estimated phase function in the frequency range from 2.5 to 3.5~kHz. The zoomed-in panels show the sampled prior distributions at the initial frequency (2.5~kHz) and at 3.5~kHz, respectively. For clarity, only 50 sample points are displayed in each panel. In the figure, the parameter range of the prior distribution progressively decreases during the sampling process. By 3.5~kHz, the sample range is reduced to approximately one thirtieth of its initial extent. This shows that the prior knowledge is rapidly updated by incorporating posterior information from the preceding data point. This is exactly consistent with Bayesian learning that initial knowledge is gradually updated by incorporating the experimental data to arrive at accurate knowledge on the data values to be estimated.

One notable observation is that the estimation accuracy decreases at frequencies where the transfer function exhibits stronger fluctuations. Correspondingly, the uncertainties in Fig.~\ref{fig:Seq}(b) increase at these same frequencies. The sampling process converges rapidly after the initial step and remains confined within a narrow range above 3~kHz, indicating a stable and reliable estimation in this frequency region.

\begin{figure}[ht]
    \centering
    \includegraphics[width=0.8\linewidth]{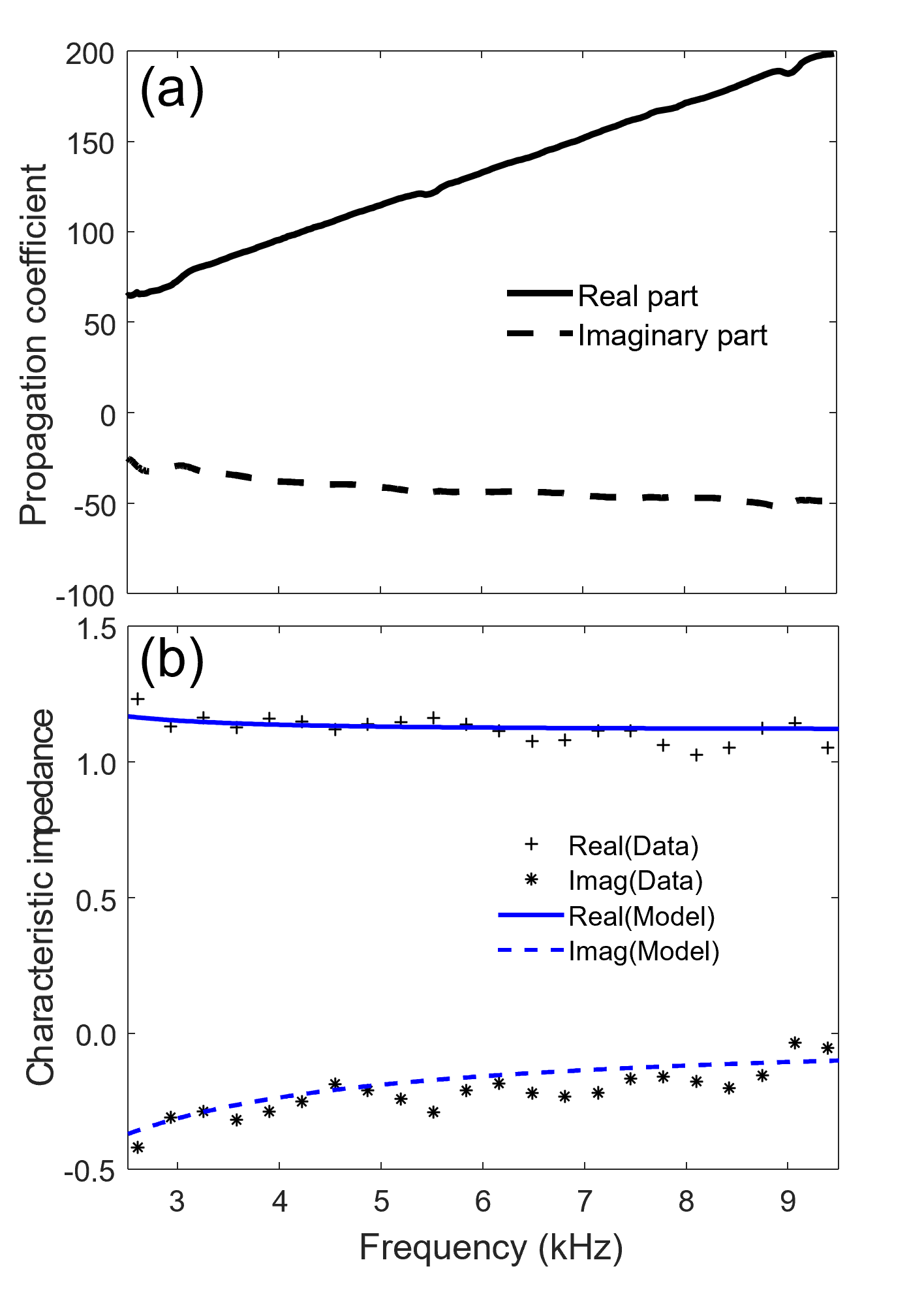}
    \caption{The estimation rresults (a) the propagation coefficient; (b) the characteristic impedance. The solid black lines are the real parts, while the dashed lines are the imaginary parts. Blue lines are calculated from the Johnson-Champoux-Allard model \citep{CA1991}.}
    \label{fig:Rst}
\end{figure}
 Figure~\ref{fig:Rst} presents the estimation results obtained from Bayesian inference. The real part of the propagation coefficient increases with frequency, as expected. The estimated characteristic impedance is compared with the classical Johnson-Champoux-Allard model~\citep{CA1991} in Fig.~\ref{fig:Rst}(b). One noticeable feature is the presence of ripples in the estimated parameters over the high frequency range. These ripples may be partly attributed to imperfections in microphone mounting. Although efforts were made to mount the 1/4~inch microphone flush with the tube wall, small surface roughness between the microphone and the tube wall may still perturb the sound field inside the tube at high frequencies. Another possible source of error is the imperfect cut of the testing sample. The roughness of the edge or side surfaces may prevent a perfect fit with the tube, leading to additional measurement uncertainty. Furthermore, the ill-posedness of the testing sample~\citep{Roncen2022} may also contribute to the observed high-frequency ripples.

One possible solution is to use a square tube instead of a circular tube for multiple-microphone measurement. 
Mounting microphones flush with the tube wall is generally easier in a square tube, which may help reduce mounting-related disturbances and improve measurement accuracy.

\section{Discussion}\label{sec:discuss}

\begin{figure}[t]
    \centering
    \includegraphics[width=0.8\linewidth]{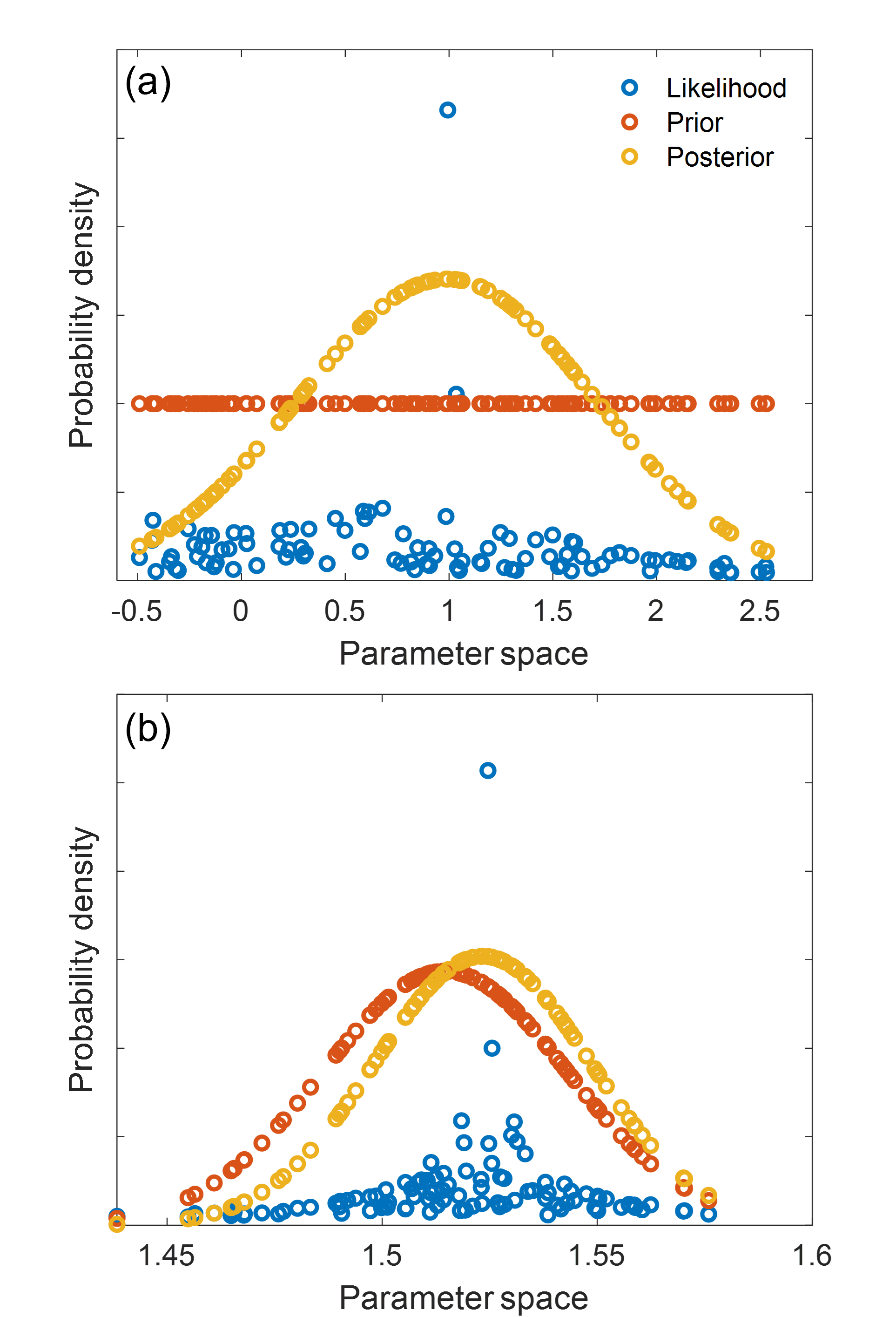}
    \caption{Sampled distributions for the real part of the phase function $\underline \theta$ at (a) the initial step; (b) the sequential phase (at 3.5 kHz).}
    \label{fig:Smppdf}
\end{figure}
Figure~\ref{fig:Smppdf} graphically illustrates the Bayes’ theorem using scatter plots of the sampled distributions. Even when a uniform prior is assigned, the likelihood function exhibits a pronounced peak. The prior at each data point is taken as the posterior from the previous data point. Note that this differs from using the posterior from a previous global estimation as the prior. Since each data point of the phase-function is estimated independently, the data $\underline{H}_{d0}$ effectively changes throughout the sequential inference process.  

Figure~\ref{fig:Smppdf}(a) illustrates how the prior distribution is updated at the the inital step (at 2.5~kHz) upon assignment of a uniform prior distribution, while 
Fig.~\ref{fig:Smppdf}(b) illustrates how the prior distribution is updated at one data point (at 3.5~kHz) by assigning a Gaussian prior to arrive at the posterior.The mean and the variance of the Gaussian assignment are informed by the posterior distribution from the previous data point 
Note that the posterior distribution is shifted relative to the prior, reflecting the strong influence of the observed data through the likelihood. In comparison with the prior and posterior distributions at the initial step, the distributions at 3~kHz are confined to a much narrower range. 

The likelihood is in the form of a Student T-distribution as in Eq.~\eqref{eq:student}. This assignment leads to the number of data points $K = 1$. Consequently, the likelihood function is inversely proportional to the residual error. Even then, Bayesian theory can still provide reliable probabilistic inference based on a data point within one frequency bin. When the residual error approaches zero, the likelihood becomes infinitely large. Therefore, the likelihood can be viewed as a unit sample sequence. 

Bayes' theorem Eq. \eqref{eq:bayes} represents how one's prior knowledge is updated in the presence of the data given the model. This Bayesian updating process is expressed as the posterior being proportional to the multiplication of two probability densities, the prior and the likelihood function. This probabilistic calculation essentially implements a sampling of the likelihood by assigning an appropriate prior distribution density; in this context, the Importance Sampling~\citep{kloek1978importancesampling} is employed. A set of Gaussian (or initially a uniform) distributed parameter samples enters into the likelihood to arrive at the posterior samples. In essence, this updating process as shown in Fig.~\ref{fig:Smppdf} is for the extremely sharply peaked likelihood to modify the prior probability/knowledge of the parameter to be estimated. Since the likelihood is so sharp as to be viewed as a unit sample sequence~\citep{openheim1989} over the parameter space, the posterior becomes peaked exactly at the position where the likelihood manifests itself as a sharp peak. Figure~\ref{fig:Seq}~(c) shows a set of  random parameter values to sample it. 

\begin{figure}[ht]
    \centering
    \includegraphics[width=0.7\linewidth]{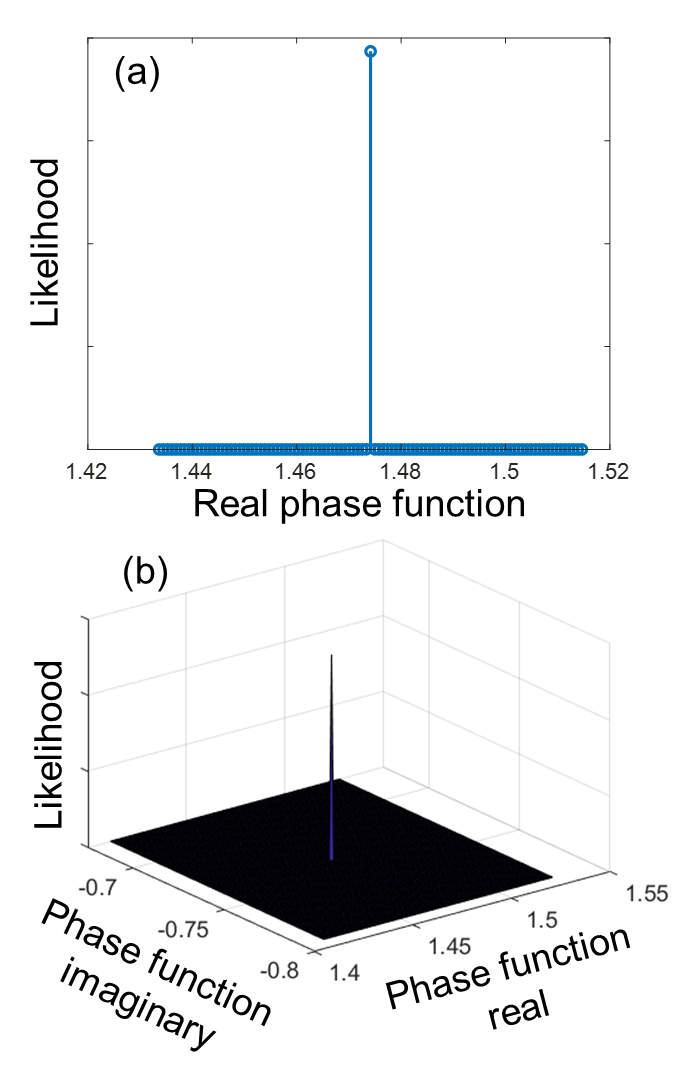}
    \caption{Sharply peaked likelihood function over the parameter space of one set of experimental data. The probability indicates that an extremely narrow region is the highest degree of confidence that the parameters to be estimated fall into this region.}
    \label{fig:pdf}
\end{figure}
Using experimentally measured transfer function data, Figure~\ref{fig:pdf}~(a) shows the sharply peaked likelihood function for the real part of the propagation coefficient, while Figure~\ref{fig:pdf}(b) illustrates the likelihood distribution over the entire complex plane. A single value is strongly favored, with the probability rapidly diminishing elsewhere. The resulting likelihood distribution closely resembles a unit-sample sequence, since the likelihood becomes unbounded when the residual error approaches zero. Consequently, when the model closely matches the data, the likelihood exhibits a pronounced peak.

In this work, model-based Bayesian inference is applied sequentially to estimate the unwrapped propagation coefficient. Bayes’ theorem represents how one’s belief in the hypothesis (the model) is updated in light of the data~\citep{xiang2020}.  
It describes how prior knowledge is updated and refined in the presence of observed data. Even a single observed data point can be used to update prior knowledge for subsequent prediction or estimation. The background information $I$ encompasses the experimenter's physical understanding of both the model and the data prior to observation. 

\begin{figure}[ht]
    \centerline{
    \includegraphics[width=0.8\linewidth]{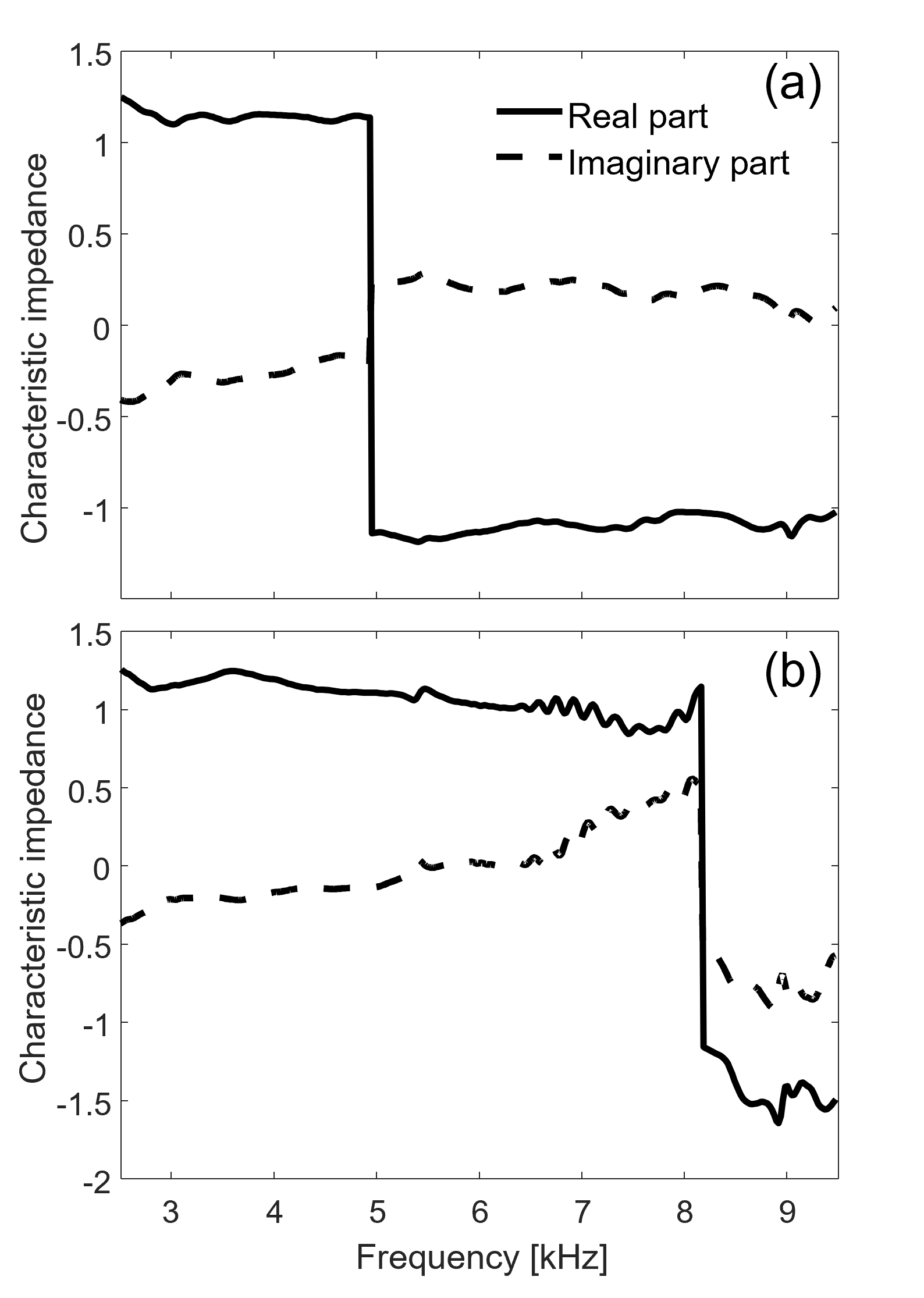}}
    \caption{Discontinuity in the measured characteristic impedance of two different materials (both are made from melamine foam) with a thickness of (a) 27.5 mm; (b) 16.0 mm.}
    \label{fig:jump}
\end{figure}
Figure~\ref{fig:jump} illustrates two experimental results for showcasing the discontinuities observed in the characteristic impedance. Shifts are evident in both the real and imaginary parts of the characteristic impedance across the frequency domain. The same material sample is measured with two different thicknesses: the testing material in Fig.~\ref{fig:jump}(a) has a thickness of 27.5~mm, whereas that in Fig.~\ref{fig:jump}(b) has a thickness of 16~mm. The frequency at which the phase jump occurs in Fig.~\ref{fig:jump}(a) is lower than that in Fig.~\ref{fig:jump}(b), which is consistent with the calculation shown in Fig.~\ref{fig:ThkScale}.

According to cylindrical mode decomposition, multiple impulse responses at Mic~0 are required for the multiple-microphone method. To preserve the rigidity of the back termination, only a single hole is drilled in the metal block, located off-center. During the measurements, the metal block is rotated by $90^\circ$ in four successive orientations to acquire impulse responses corresponding to all four microphone positions flush with the rigid backing. In this manner, the four impulse responses required for cylindrical mode decomposition are obtained while minimizing any degradation of the back termination rigidity.

In addition to the three-microphone method, the four-microphone method by Song and Bolton \citep{song2000} will similarly encounter discontinuity problems within broad frequency ranges, to determine the propagation coefficient and the characteristic impedance of porous media under test. There the necessary steps require inversion of cosine and sine functions as well. Any transfer function method involving the inversion of trigonometric functions can benefit from this approach.

\section{Conclusion}\label{sec:conc}
This work presents a robust Bayesian framework for characterizing porous materials using a multiple-microphone three-microphone impedance tube method over extended frequency ranges. By incorporating cylindrical mode decomposition and circumferential averaging of multiple microphones, the valid frequency range of characteristic impedance measurements is nearly doubled without reducing tube diameter or sample size.

A key challenge encountered in broadband measurements is the appearance of undesirable discontinuities in the measured propagation coefficient and characteristic impedance. These discontinuities arise from the intrinsic multivalued nature of the $\arccos(\cdot)$ function used to invert the measured transfer function. Unlike the well-studied phase unwrapping of the $\arctan(\cdot)$ function, phase unwrapping for $\arccos(\cdot)$ lacks sufficient analytical information to determine the correct branch of the complex phase function.

To address this issue, Bayesian inference is applied in a sequential manner to unwrap the complex-valued phase function, from which the propagation coefficient and the characteristic impedance are obtained. By treating each frequency bin as an independent inference problem and propagating the posterior information on the current data point forward as prior knowledge of the subsequent data point, the proposed approach adaptively resolves phase wrapping without requiring prior knowledge of the wrapping frequency. The inferred parameters closely match the experimental data and remain physically consistent across the extended frequency range.

The results demonstrate that Bayesian inference provides a powerful alternative to analytical phase-unwrapping methods, particularly in cases where inverse trigonometric functions introduce ambiguity. Although demonstrated here for a three-microphone impedance tube, the proposed framework is general and can be readily applied to other tube-based characterization methods that involve inversion of trigonometric transfer-function relationships.

Future work will explore alternative tube geometries, such as square cross-sections, to further extend the usable frequency range and reduce mounting-related uncertainties. The Bayesian framework presented in this study offers a principled and flexible foundation for broadband acoustic material characterization in the presence of phase ambiguity.

\section*{AUTHOR DECLARATIONS}
\subsection*{Conflict of interest}
    All authors have no conflicts of interest to declare.

\section*{Data Avaliability}
The data that support the findings of this study are available from the corresponding author upon reasonable request.

\bibliography{Ziqibib.bib}

@article{chen2022bayesian,
  title={Bayesian estimations of dissipation, sound speed, and microphone positions in impedance tubes},
  author={Chen, Ziqi and Xiang, Ning and Fackler, Cameron J},
  journal={JASA Express Letters},
  volume={2},
  number={8},
  year={2022},
  publisher={AIP Publishing}
}

@article{kloek1978importancesampling,
  title={Bayesian Estimates of Equation System Parameters: An Application of Integration by Monte Carlo},
  author={T. Kloek and  H. K. {van Dijk}},
  journal={Econometrica},
  volume={46},
  number={1},
  pages={1-19},
  year={1978},
  publisher={ The Econometric Society}
}

@article{salissou2010,
  title={Wideband characterization of the complex wave number and characteristic impedance 
            of sound absorbers},
  author={Yacoubou Salissou and Raymond Panneton},
  journal={J. Acoust. Soc. Am.},
  volume={128},
  number={5},
  pages={2868-2876},
  year={2010},
  publisher={Acoustical Society of America}
}

@inproceedings{panzer2019extracting,
  title={Extracting the Fundamental Mode from Sound Pressure Measurements in an Acoustic Tube},
  author={Panzer, Joerg},
  booktitle={Proc. AES 157th Conv. 147},
  year={2019},
  organization={Audio Engineering Society}
}

@article{chen2024bayesian,
  title={Bayesian estimation of dissipation and sound speed in tube measurements using a transfer-function model},
  author={Chen, Ziqi and Xiang, Ning},
  journal={J. Acoust. Soc. Am.},
  volume={155},
  number={4},
  pages={2646--2658},
  year={2024},
  publisher={AIP Publishing}
}

@misc{ASTM1050,
author={{ASTM E1050-19}},
year={2019},
title={Standard {Test} {Method} for {Impedance} and {Absorption} of {Acoustical Materials Using} a
{Tube}, {Two Microphones} and a {Digital Frequency Analysis System}},
note={(American Society for Testing and Materials, Philadelphia)}
}

@article{Roncen2022,
  title={Addressing the ill-posedness of multi-layer porous media   
    characterization in impedance tubes through the addition of air
    gaps behind the sample: Numerical validation},
  author={Rémi Roncen and Zine El Abiddine Fellah and Erick Ogam},
  journal={J. Sound \& Vib. },
  volume={520},
  number={},
  pages={116601},
  year={2022}
}

@article{sanada2018extension,
  title={Extension of the frequency range of normal-incidence sound absorption coefficient measurement in impedance tube using four or eight microphones},
  author={Sanada, Akira and Iwata, Kazuhiro and Nakagawa, Hiroshi},
  journal={Acoust. Sci. Technol.},
  volume={39},
  number={5},
  pages={335--342},
  year={2018},
  publisher={ACOUSTICAL SOCIETY OF JAPAN}
}

@inproceedings{iwase1998new,
  title={A new measuring method for sound propagation constant by using sound tube without any air spaces back of a test material},
  author={Iwase, T and Izumi, Y and Kawabata, R},
  booktitle={INTER-NOISE Proc.},
  volume={1998},
  number={4},
  pages={1265--1268},
  year={1998},
  organization={Institute of Noise Control Engineering}
}

@article{eser2023free,
  title={Free-field characterization of locally reacting sound absorbers using {B}ayesian inference with sequential frequency transfer},
  author={Eser, M and Mannhardt, S and Gurbuz, C and Brandao, E and Marburg, S},
  journal={Mech. Syst. Signal Process.},
  volume={205},
  pages={110780},
  year={2023},
  publisher={Elsevier}
}

@article{xiang2020,
  title={{Model-based Bayesian analysis in acoustics—A tutorial}},
  author={Xiang, Ning},
  journal={J. Acoust. Soc. Am.},
  volume={148},
  number={2},
  pages={1101-1120},
  year={2020},
  publisher={Acoustical Society of America}
}

@book{xiang2021,
    author = {Ning Xiang and Jens Blauert},
    title = {Acoustics for Engineers -- Troy Lectures},
    publisher =  {Springer-Verlag},
    edition ={3rd.},
    address = {Berlin Heidelberg},
    note = {{Chap.8}},
    year = 2021 
}

@article{yang2025data,
  title={Data-driven impedance tube method for prediction of normal sound absorption coefficient},
  author={Yang, Zu-Jie and Zhang, Yong-Bin and Xu, Liang and Zhang, Xiao-Zheng and Bi, Chuan-Xing},
  journal={J. Acoust. Soc. Am.},
  volume={157},
  number={4},
  pages={2422--2432},
  year={2025},
  publisher={AIP Publishing}
}

@article{rienstra2015fundamentals,
  title={Fundamentals of duct acoustics},
  author={Rienstra, Sjoerd W},
  journal={Von Karman Institute Lecture Notes},
  volume={598},
  year={2015},
  publisher={Von Karman Institute Sint-Genesius-Rode, Belgium}
}

@misc{ISO10534,
author = {{ISO 10534–2}},
booktitle = {International Standards Organization},
pages = {149--165},
title = {{Acoustics–Determination sound absorption coefficient and impedance in impedance tubes. 2. Transfer function method}},
volume = {},
note ={(Geneva)},
year = {1998}
}

@article{chu1986,
author = {W. T. Chu},
journal = {J. Acoust. Soc. Am.},
pages = {347--348},
title = {Extension of the two-microphone transfer function method for impedance tube measurements},
volume = {80},
number = {1},
year = {1986}
}

@article{song2000,
  title = {A transfer-matrix approach for estimating the characteristic
            impedance and wave numbers of limp and rigid porous materials},
  author = {Bryan H. Song and J. Stuart Bolton},
  journal = {J. Acoust. Soc. Am.},
  year = {2000},
  volume={107},
  number={3},
  pages={1131-1152},
  publisher={Acoustical Society of America}
}

@ARTICLE{Tribolet1977,
  author={Tribolet, J.},
  journal={IEEE Trans. Acoust., Speech, Signal Process.}, 
  title={A new phase unwrapping algorithm}, 
  year={1977},
  volume={25},
  number={2},
  pages={170-177},
  doi={10.1109/TASSP.1977.1162923}}

@article{ABOM1989,
title = {Modal decomposition in ducts based on transfer function measurements between microphone pairs},
journal = {J. Sound Vib. },
volume = {135},
number = {1},
pages = {95-114},
year = {1989},
issn = {0022-460X},
doi = {https://doi.org/10.1016/0022-460X(89)90757-8},
url = {https://www.sciencedirect.com/science/article/pii/0022460X89907578},
author = {M. Åbom}
}

@inproceedings{cuenca2013angular,
  title={Angular and radial acoustic mode detection in large cylindrical ducts by sequential measurements using a single ring of microphones},
  author={Cuenca, Jacques and Hallez, Rapha{\"e}l and Peeters, Bart},
  booktitle={20th International Congress on Sound and Vibration 2013, ICSV 2013},
  volume={1},
  pages={721--728},
  year={2013}
}

@article{CA1991,
    author = {Champoux, Yvan and Allard, Jean‐F.},
    title = {Dynamic tortuosity and bulk modulus in air‐saturated porous media},
    journal = {Journal of Applied Physics},
    volume = {70},
    number = {4},
    pages = {1975-1979},
    year = {1991},
    month = {08},
    issn = {0021-8979},
    doi = {10.1063/1.349482},
    url = {https://doi.org/10.1063/1.349482},
    eprint = {https://pubs.aip.org/aip/jap/article-pdf/70/4/1975/18644305/1975_1_online.pdf},
}

@article{Chung1980a,
author = {J. Y. Chung and D. A. Blaser},
journal = {J. Acoust. Soc. Am.},
pages = {907-913},
title = {Transfer function method of measuring in-duct acoustic properties, {I. Theory}},
volume = {68},
number = {},
year = {1980}
}

@book{openheim1989,
    author = {Alan V. Oppenheim and Ronald W. Schafer},
    title = {Discrete-Time Signal Processing},
    publisher =  {Prentice-hall},
    edition ={1st},
    note = {{Signal Processing Series}},
    year = 1989 
}

@article{JKD1987, 
    title={Theory of dynamic permeability and tortuosity in fluid-saturated porous media}, 
    volume={176}, DOI={10.1017/S0022112087000727}, 
    journal={Journal of Fluid Mechanics}, 
    author={Johnson, David Linton and Koplik, Joel and Dashen, Roger}, 
    year={1987}, 
    pages={379–402}}

@article{CA1992,
    author = {Allard, Jean‐F. and Champoux, Yvan},
    title = {New empirical equations for sound propagation in rigid frame fibrous materials},
    journal = {J. Acoust. Soc. Am.},
    volume = {91},
    number = {6},
    pages = {3346-3353},
    year = {1992},
    month = {06},
    issn = {0001-4966},
    doi = {10.1121/1.402824},
    url = {https://doi.org/10.1121/1.402824},
    eprint = {https://pubs.aip.org/asa/jasa/article-pdf/91/6/3346/11806860/3346_1_online.pdf},
}

\end{document}